\documentclass[conference]{IEEEtran}
\usepackage[T1]{fontenc}

\usepackage[ruled]{algorithm2e}  
\usepackage{amsmath}
\usepackage{amssymb}
\usepackage{cite}
\usepackage{color}
\usepackage{courier}
\usepackage{./sty/easygraphic}
\usepackage{etextools}
\usepackage{etoolbox}
\usepackage{float}
\usepackage[keeplastbox]{flushend}
\usepackage{hyperref}
\usepackage{listings}
\usepackage{mathtools}
\usepackage{pdfpages}
\usepackage{pgfkeys}
\usepackage{pifont}
\usepackage[binary-units=true]{siunitx}
\usepackage{subcaption}
\usepackage{tabu}
\usepackage{todonotes}
\usepackage{xparse}

\usepackage[capitalise]{cleveref}

\let\todoold\todo
\renewcommand{\todo}[1]{\todoold[inline,caption={2do}]{\begin{minipage}{\textwidth-4pt}#1\end{minipage}}}

\def\figureFontSizeName{footnotesize}
\def\figureFontSize{\csname \figureFontSizeName \endcsname}
\SetAlCapFnt{\rm\figureFontSize}


\def\etal{\mbox{\em et al.}}
\newcommand\bigO[1]{$\mathcal{O}(#1)$}  
\def\bigOn{\bigO{N}}

\def\bigOnn{\bigO{N^2}}
\def\bigOnsqrtn{\bigO{N\sqrt{N}}}

\DeclareDocumentCommand{\hanging}{O{1em} O{0em} +m}{%
    \begin{list}
            {}
            {\setlength{\itemindent}{-#1}%
                \setlength{\leftmargin}{#2+#1}%
                \setlength{\itemsep}{0pt}%
                \setlength{\parsep}{\parskip}%
                \setlength{\topsep}{\parskip}%
            }
    \setlength{\parindent}{-#1}%
    \item[]
    #3
    \end{list}
}

\pgfkeys{/fig/.is family, /fig,
    pos/.estore in=\figPos,
    label/.estore in=\figLabel,
    lof/.estore in=\figLof,
    default/.style = {
        pos=t,
        label=,
        lof=,
    }
}
\DeclareDocumentCommand{\fig}{O{} +m +m}{%
    \pgfkeys{/fig, default, #1}

    \edef\figPoss{[\figPos]}
    \expandaftercmds{\begin{figure}}{\figPoss}
        \centering
        {
            \figureFontSize
            #2
        }
        \ifdefempty{\figLof}{%
            \caption{#3}}{%
            \caption[\figLof]{#3}}
        \ifdefempty{\figLabel}{}{\label{\figLabel}}
    \end{figure}
}

\pgfkeys{/subfig/.is family, /subfig,
    caption/.estore in=\subfigCaption,
    label/.estore in=\subfigLabel,
    width/.estore in=\subfigWidth,
    default/.style = {
        caption={},  
        width=,
        label=,
    }
}
\DeclareDocumentCommand{\subfig}{O{} +m}{%
    \pgfkeys{/subfig, default, #1}
    \begin{subfigure}[t]{\ifdefempty{\subfigWidth}{\linewidth}{\subfigWidth}}
        \centering
        #2
        \caption{\subfigCaption}
        \ifdefempty{\subfigLabel}{}{\label{\subfigLabel}}
    \end{subfigure}%
}

\pgfkeys{/alg/.is family, /alg,
    label/.estore in=\algLabel,
    pos/.estore in=\algPos,
    default/.style = {
        label=,
        pos=t,
    }
}
\makeatletter
\newcommand{\algRemovelatexerror}{\let\@oldlatex@error\@latex@error \let\@latex@error\@gobble}
\newcommand{\algRestorelatexerror}{\let\@latex@error\@oldlatex@error}
\makeatother
\newlength\algboldlen
\DeclareDocumentCommand{\alg}{O{} +m +m}{%
    \pgfkeys{/alg, default, #1}

    \edef\algPoss{[\algPos]}
    \expandaftercmds{\begin{figure}}{\algPoss}
        \centering
        \algRemovelatexerror \begin{algorithm}[H] \algRestorelatexerror
            \caption{\figureFontSize #2}
            \ifdefempty{\algLabel}{}{\label{\algLabel}}

            \newcommand\BoldKw[2]{
                \settowidth\algboldlen{{\bf ##1: }}
                \begin{minipage}{\linewidth-\leftmargin+\itemindent}%
                    {\bf ##1:} \hanging[1.5em][\algboldlen]{\vspace{-1em} ##2}
                \end{minipage}}
            \newcommand\Input[1]{\BoldKw{Input}{##1}}
            \newcommand\Output[1]{\BoldKw{Output}{##1}}

            \figureFontSize #3
        \end{algorithm}
    \end{figure}
}

\pgfkeys{/wtable/.is family, /wtable,
    padh/.estore in=\wtablePadh,
    pos/.estore in=\wtablePos,
    label/.estore in=\wtableLabel,
    lof/.estore in=\wtableLof,
    default/.style = {
        pos=t,
        label=,
        lof=,
        padh=,
    }
}
\DeclareDocumentCommand{\wtable}{O{} +m +m +m}{%
    \pgfkeys{/wtable, default, #1}

    \newlength{\savedPadh}
    \setlength{\savedPadh}{\tabcolsep}
    \ifdefempty{\wtablePadh}{}{%
        \setlength{\tabcolsep}{\wtablePadh}
    }

    \edef\wtablePoss{[\wtablePos]}
    \expandaftercmds{\begin{table}}{\wtablePoss}
        \centering
        \ifdefempty{\wtableLof}{%
            \caption{#2}}{%
            \caption[\wtableLof]{#2}}

        \begin{tabu} to 1.00\linewidth {#3}
            \hline
            #4
            \hline
        \end{tabu}

        \ifdefempty{\wtableLabel}{}{\label{\wtableLabel}}
    \end{table}

    \setlength{\tabcolsep}{\savedPadh}
}

\author{%
    Walt Woods and Christof Teuscher \\
    Department of Electrical and Computer Engineering \\
    Portland State University, Portland, OR, USA \\
    \{wwoods, teuscher\}@pdx.edu%
    \vspace{-2em}}

\title{Fast and Accurate Sparse Coding of Visual Stimuli with a Simple, Ultra-Low-Energy Spiking Architecture \vspace{-0.5em}}

\begin{document}
\maketitle

\begin{abstract}%
    Memristive crossbars have become a popular means for realizing unsupervised and supervised learning techniques.  In previous neuromorphic architectures with leaky integrate-and-fire neurons, the crossbar itself has been separated from the neuron capacitors to preserve mathematical rigor.  In this work, we sought to design a simplified sparse coding circuit without this restriction, resulting in a fast circuit that approximated a sparse coding operation at a minimal loss in accuracy.  We showed that connecting the neurons directly to the crossbar resulted in a more energy-efficient sparse coding architecture, and alleviated the need to pre-normalize receptive fields.  This work provides derivations for the design of such a network, named the Simple Spiking Locally Competitive Algorithm, or SSLCA, as well as CMOS designs and results on the CIFAR and MNIST datasets.  Compared to a non-spiking, non-approximate model which scored \SI{33}{\%} on CIFAR-10 with a single-layer classifier, this hardware scored \SI{32}{\%} accuracy.  When used with a state-of-the-art deep learning classifier, the non-spiking model achieved \SI{82}{\%} and our simplified, spiking model achieved \SI{80}{\%}, while compressing the input data by \mbox{\SI{92}{\%}}.  Compared to a previously proposed spiking model, our proposed hardware consumed \SI{99}{\%} less energy to do the same work at \SI{21}{\times} the throughput.  Accuracy held out with online learning to a write variance of \SI{3}{\%}, suitable for the often-reported 4-bit resolution required for neuromorphic algorithms; with offline learning to a write variance of \protect\SI{27}{\%}; and with read variance to \SI{40}{\%}.  The proposed architecture's excellent accuracy, throughput, and significantly lower energy usage demonstrate the utility of our innovations.
\end{abstract}

\begin{IEEEkeywords}
sparse coding, locally competitive algorithm, memristors, neuromorphic architecture, spiking architecture
\end{IEEEkeywords}

\vspace{-1em}


\section{Introduction}\label{sec:intro}


%
\newcommand{\KnagLowPower}{\SI{48}{\pico\joule\per input}}
\newcommand{\KnagLowThroughput}{\SI{0.55}{\mega Ops\per\second}}
\newcommand{\KnagHighPower}{\SI{176}{\pico\joule\per input}}
\newcommand{\KnagHighThroughput}{\SI{4.8}{\mega Ops\per\second}}
\newcommand{\KnagHighThroughputVsShapero}{\SI{120}{\times}}

\newcommand{\KimHighPower}{\SI{26.4}{\pico\joule\per input}}
\newcommand{\KimHighThroughput}{\SI{9.9}{\mega Ops\per\second}}

\newcommand{\ShaperoThroughput}{\SI{40}{\kilo Ops\per\second}}

\newcommand{\ThisCifarEightPower}{\SI{1.77}{\pico\joule\per input}}
\newcommand{\ThisCifarEightThroughput}{\SI{100}{\mega Ops\per\second}}

\newcommand{\ThisMnistPower}{\SI{0.26}{\pico\joule\per input}}
\newcommand{\ThisMnistThroughput}{\SI{100}{\mega Ops\per\second}}

\newcommand{\ThisKnagCifarLowPower}{\SI{96}{\percent}}
\newcommand{\ThisKnagCifarLowThroughput}{\SI{180}{\times}}
\newcommand{\ThisKnagCifarHighPower}{\SI{99}{\percent}}
\newcommand{\ThisKnagCifarHighThroughput}{\SI{21}{\times}}

Sparse coding, accomplished through algorithms that encode an input stimulus in a new basis with few non-zero elements, has been shown to improve image classification accuracy with single-layer classifiers \cite{Coates2011}.  These algorithms have also been shown to reduce the learning time required for backpropagation \cite{Ammar2012}.  Since there are few non-zero elements, sparse coding also provides a means for minimizing the bandwidth required to transfer sensor data amongst multiple processors, or to store that data in long-term storage.  In recent years sparse coding has gained additional traction as a result of biological evidence that the V1 visual layer in mammalian cortices performs similar functionality \cite{Masquelier2007,Zamarreno-Ramos2011,Zhu2015}.  The revelation that sparse coding algorithms should be part of a neuromorphic learning system has intensified research using these algorithms.  The implementation of neuromorphic algorithms on custom {\em Application-Specific Integrated Circuits} (ASICs) has also been a wildly popular area of study largely due to the development of programmable, variable-resistance nanodevices, named memristors, that can be used to realize the synapses needed in a more compact, energy efficient form \cite{Querlioz2012,Zamarreno-Ramos2011,Jo2010,Masquelier2007,Woods2015tnano,Payvand2015,Bennett2015,Olshausen2017}.


Our work generally extends from the {\em Locally Competitive Algorithm} (LCA) proposed by Rozell \etal\ in 2008, an optimal solver for the sparse coding problem \cite{Rozell2008}, coupled with Oja's rule, used to repeatedly tune the dictionary towards an optimal solution for a set of training inputs without requiring a supervisory signal \cite{Oja1982}.  The LCA was chosen as the most promising sparse coding algorithm due to its use of inhibitory forces to force an optimally sparse and stable solution.  We give special consideration to the case where the LCA is used as a front end to a traditional supervised classifier.  We investigate the efficacy of the supervised classifier to classify an input stimulus as either the value of a handwritten digit or the type of object in a tiny image:  MNIST and CIFAR-10, respectively \cite{Lecun1998,Krizhevsky2009}.  This approach leverages the transformation of input data from its native space to a decorrelated space via the LCA, and then uses traditional machine learning techniques to classify the stimulus based on its decorrelated representation.  We have used this approach in the past \cite{Woods2015tnano}.  In practice, there are several benefits to this approach: improved accuracy due to the stability of the decorrelated representation (an effect similar to dropout), and the ability to compress the input stimulus between the measuring device and the identification layer.  A single frame of HD video data consists of approximately \SI{6.2}{\mega\bit} of information, which the LCA could compress down to \mbox{\SI{1.0}{\mega\bit}}, an \mbox{\SI{84}{\%}} reduction, at a \mbox{\em Root-Mean-Square Error} (RMSE) of \mbox{\SI{5.8}{\%}}, or down to \mbox{\SI{0.6}{\mega\bit}}, a \mbox{\SI{90}{\%}} reduction, at an RMSE of \mbox{\SI{7.0}{\%}}.  For video surveillance systems or autonomously driving vehicles, this means that several cameras could be wired into a single, high-speed, low-energy sparse coding device, greatly reducing the needed communication bandwidth for the system.

In this work we set out to provide a low-power, hardware-friendly realization of an LCA-like algorithm using a spiking framework.  To maximize power savings, the model was simplified and the drawbacks of that simplification were investigated.  Spikes were utilized to save power; the efficacy of spikes for power saving was explored.  The proposed architecture has been named the {\em Simple Spiking Locally Competitive Algorithm} (SSLCA).  The SSLCA was compared with the original LCA's ODE \cite{Rozell2008} as well as their later spiking work in Shapero \etal\ \cite{Shapero2013}.  Both the MNIST and CIFAR-10 datasets were used for the comparison.  Through these comparisons, we found that our proposal demonstrated excellent power and scaling qualities.  It is our hope that this work provides the basis for efficient, next-generation sparse-coding hardware.

\section{Related Work}\label{sec:related}

The original work on LCAs was Rozell \etal, 2008 \cite{Rozell2008}.  Rozell \etal\ sought to improve upon prior sparse coding algorithms by deriving an optimal expression to both minimize the sparse coding equation (\cref{eq:sparse}) and smooth the generated sparse representation when given time-varying input.  Their work derived an {\em Ordinary Differential Equation} (ODE) that solved both of these problems (\cref{eq:lca}).  By providing competition amongst outputs, the number of active outputs may be minimized while simultaneously maximizing the fidelity of the reconstruction produced.  The sparse coding equation that is minimized by the LCA, with the input stimulus denoted as $\mathbf{s}$, the sparse code of coefficients to reconstruct the input as $\mathbf{a}$, the bases that comprise the sparse code as the matrix $\mathbf{\Phi}$, the reconstruction of the input as $\mathbf{\hat{s}} = \mathbf{\Phi}\mathbf{a}$, and a cost function $\lambda C(\cdot)$ that expresses the trade-off between reconstruction quality and the sparsity of the final solution $\mathbf{a}$, is written as:

\begin{align}
    E(t) &= \frac{1}{2}||\mathbf{s}(t) - \mathbf{\hat{s}}(t)||^2 + \lambda \sum_m C(\mathbf{a}_m(t)). \label{eq:sparse}
\end{align}

$E(t)$ is minimized by descending an ODE derived by Rozell \etal\ \cite{Rozell2008} to its natural steady state:

\begin{align}
    \mathbf{a}_m(t) &= T_{\lambda}(\mathbf{u}_m(t)), \nonumber \\
    \mathbf{\dot{u}}_m(t) &= \frac{1}{\tau}\left[ \mathbf{b}_m(t) - \mathbf{u}_m(t) - \sum_{n\ne m} \mathbf{G}_{m,n}\mathbf{a}_n(t) \right], \label{eq:lca}
\end{align}

\noindent where $\mathbf{u_m}$ is an underlying, non-sparse state for the $m$th neuron that is thresholded with a function $T_{\lambda}(\cdot)$ to produce the resulting sparse code element $\mathbf{a_m}$, $\mathbf{b_m}$ is the inner product of the $m$th neuron's receptive field and the input, and $\mathbf{G}_{m,n}$ is the inner product of the $m$th and $n$th receptive fields, minus 1 if $m=n$.  Approximating the integral of \mbox{\cref{eq:lca}} with sufficiently small step sizes will always end in a stable $\mathbf{u}$, corresponding to a generated sparse code $\mathbf{a}$.  See the original LCA paper for more details \mbox{\cite{Rozell2008}}.

\fig[label=fig:related:compare]
    {\graphic{plots/compare-tech-power.pdf}}
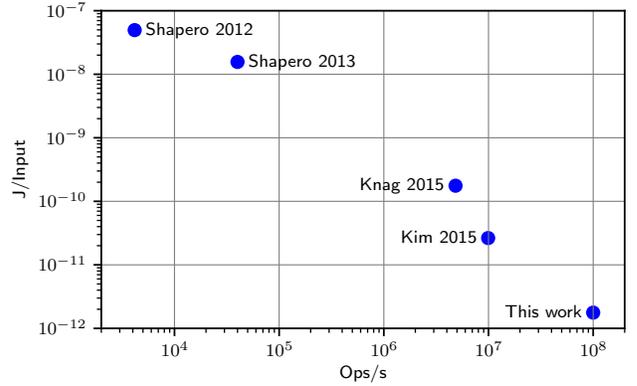
    {Comparison of the SSLCA's energy efficiency and throughput, presented in this work, with previous state-of-the-art results.  One ``Op,'' or operation, is the complete generation of a sparse code from a single set of inputs.}

Due to the popularity of neuromorphic algorithms, there has been a lot of prior work relating to ASICs for neuromorphic sparse coding architectures, both within C. Rozell's group and other research groups.  The relative performance of these in terms of energy efficiency and throughput are shown in \cref{fig:related:compare}.  Note that the throughput axis is presented in ``Ops/s'' as opposed to ``Inputs/s,'' so that the scale of an architecture has no effect on the measures reported.  For example, if an architecture were built to process $64$ inputs, and its performance were measured, doubling that architecture to $128$ inputs would immediately double its throughput measurement if we used ``Inputs/s'' rather than ``Ops/s.''  Using ``Ops/s'' makes the number of inputs irrelevant, which is more appropriate for architectures which have a fixed processing time regardless of the number of inputs and outputs.

Throughputs presented are inference-only; throughputs when using online training would be different in accordance with the time to set the different storage mediums used in each algorithm.  As an example in this work, many memristive models can have their resistances modified within \mbox{\SIrange{2}{20}{\nano\second}} \mbox{\cite{Woods2015}}.  If every algorithm execution led to a weight update of each device in the matrix, and each update happened serially, this would lead to e.g. \mbox{\SI{392}{\micro\second}} being added to each loop of the matrix for an example network with $784$ inputs and $50$ outputs.  However, this quantity may easily be amortized: either multiple devices might be updated at once, multiple runs of the algorithm could be integrated into each weight update, or only a portion of the weights would need to be updated each iteration.  For all neural algorithms, disabling learning results in significantly higher throughput.  Furthermore, when the problem at hand is sufficiently solved, there is no further need for the learning step, an often-used motivation for offline learning in practical applications.

While \cref{eq:lca} is an effective equation for sparse coding, prior attempts at implementing the LCA directly in hardware have suffered from a few details which prevented an efficient implementation.  For example, the LCA's matrix form can be intuitively described as reproducing the input based on a linear combination of the weight matrix columns.  The most significant term limiting the LCA's efficiency comes from the inhibition term in \cref{eq:lca}: each output column's coefficient's ODE depends on all other output columns.  For a naive implementation, and indeed the one chosen by Rozell's group in their original hardware implementation using floating gates \cite{Shapero2012}, this implies \bigOnn\ hardware scaling: doubling the number of output elements quadruples the required hardware.  While Rozell's group found that the actual power consumption scaled less quickly than the amount of hardware, at \bigOnsqrtn, the amount of hardware still scaled as \bigOnn.  Additionally, ODE convergence in their hardware was relatively slow, occurring after \SI{240}{\micro\second} \cite{Shapero2012}.  Even outside of the inhibitory term, a low-power implementation of the dot product in the ODE is non-trivial: it had to either be implemented digitally, or using next-generation components like memristors.  While variable-resistance nanodevices like memristors can compute a dot product using little power themselves, to make the computation accurate requires either a virtual ground, which consumes significant power due not only to the matching current but also due to excessive current drained through low-resistance devices \mbox{\cite{Woods2015,Woods2015tnano,Woods2017vmm}}, or requires a tuning resistance that must change based on each column's configuration \mbox{\cite{Woods2017vmm}}.  The additional power consumed by these solutions motivates the exploration of techniques that do not require the calculation of an exact dot product.

The drawbacks of power scaling and slow convergence times were addressed to some extent later by Shapero \etal\ in 2013 \cite{Shapero2013}.  That work extended the original LCA to a spiking architecture, referred to in this work as the {\em Spiking Locally Competitive Algorithm} (SLCA).  The motivation for spiking largely seems to have stemmed from biology: all biological systems appear to use spiking rather than constant signals \cite{Masquelier2007,Serrano-Gotarredona2013,Zamarreno-Ramos2011}.  Spiking models have also long been believed to consume less power, and to exhibit additional computational power due to their stochasticity \cite{Maass2015,Habenschuss2013,Hamilton2014}.  The validity of leveraging spikes to save power is discussed further in \cref{sec:model:cifar:power} of this work.  In their work, Shapero \etal\cite{Shapero2013} showed that their SLCA consumed more power than their LCA at small sizes, but that their SLCA scaled only as the desirable \bigOn, and would consume less power than the LCA at large network sizes.  Additionally, they reduced the convergence time to \SI{25}{\micro\second}, nearly \SI{90}{\%} faster than their LCA with a throughput of \ShaperoThroughput.  However, the required hardware still scaled as \bigOnn.

Other spiking networks optimized for sparse coding have been published, such as SAILnet, introduced by Zylberberg \etal\ in 2011 \cite{Zylberberg2011}.  ASICs using this architecture have been studied, with a substantial reduction in power compared to the approach presented by Shapero \etal\cite{Shapero2013}.  Knag \etal\cite{Knag2015} were capable of using the SAILnet architecture to process images using only \KnagLowPower\ for their inference logic with a throughput of \KnagLowThroughput, or using \KnagHighPower\ with a throughput of \KnagHighThroughput, \KnagHighThroughputVsShapero\ as fast as Shapero \etal\cite{Shapero2013}.  Their design was CMOS-based, and utilized a decreased resolution for weight storage: \SI{4}{} bits per excitatory or inhibitory weight.  This decision has been justified in a number of prior works dealing with how much accuracy is needed for sparse coding algorithms to perform well \cite{Pfeil2012,Woods2015tnano}.  This design was later reconfigured by Kim \mbox{\etal} \mbox{\cite{Kim2015}}.  They tested higher clock speeds, optimized the design, and generated less detailed output, resulting in a throughput of \mbox{\KimHighThroughput} at an energy efficiency of \mbox{\KimHighPower} \mbox{\cite{Kim2015}}.  However, these numbers benefit from their usage of only 256 output neurons to represent 1024 inputs, whereas Knag \mbox{\etal} used 256 output neurons to represent only 256 inputs \mbox{\cite{Kim2015,Knag2015}}.  Not only does using fewer outputs result in a worse encoding of the input, but due to the \mbox{\bigOnn} scaling properties of these networks, this also favors the power and throughput figures in \mbox{\cite{Kim2015}}.  As such, subsequent results in this work will be compared with Knag \mbox{\etal}, as their chip performs a comparable amount of work to ours \mbox{\cite{Knag2015}}.  Like the LCA, SAILnet uses a direct inhibitory weight between each pair of output neurons, yielding a scaling complexity of \bigOnn.

The closest family of algorithms that does not exhibit \bigOnn\ scaling is {\em Spike-Timing-Dependent Plasticity} (STDP).  STDP exploits what is known as ``Hebbian'' learning, where input spike events that occur at the same time as an output spike event become more likely to trigger that output spike event.  The common idiom for this behavior is, ``neurons that fire together, wire together.''  In effect, each output neuron learns to activate when a correlated set of inputs fires together.  This is very similar to what happens in sparse coding, where a neuron responds to a specific pattern in the input.  The primary differences are that STDP makes no effort to preserve the information found in the input and STDP does not implement inhibition amongst neurons.  Rather, the purpose of STDP is to flag which features are present in the input and how prevalent they are, without regard for the other features present.  Sparse coding, on the other hand, will suppress output of a feature that is already represented by a combination of other features.  Both techniques are a form of unsupervised learning, except sparse coding requires some inhibitory terms while STDP does not.  This gives STDP the desirable quality of \bigOn\ scaling.  Due to its excellent scaling properties, STDP was used in one of the earliest attempts to replicate the features found in mammalian visual cortices \cite{Masquelier2007}, has been explored as an autoencoder \cite{Burbank2015}, and has been used to generate unsupervised features for digit classification on the MNIST digit database \cite{Querlioz2012}.  STDP is also one of the dominant architectures researched using next-generation nanodevices such as memristors \cite{Payvand2015,Querlioz2012,Bennett2015,Serrano-Gotarredona2013,Jo2010,Zamarreno-Ramos2011}.  The downside of an STDP approach to input encoding is that more output neurons are required due to the lack of inhibition; with \SI{50}{} neurons, prior research showed that STDP achieved 80\% accuracy on MNIST, while a sparse coding layer using LCA achieved 85\% \cite{Querlioz2012,Woods2015tnano}.  If a sparse coding algorithm were implemented with the same efficiency as STDP, it would be the preferred method of unsupervised training, as it conveys more depth of information.  That is what our work set out to accomplish: to close the gap between STDP and sparse coding algorithms such that there exists a simple and efficient algorithm for sparse coding.

Recent work by Sheridan \etal\ showed that their group has manufactured memristive crossbars and applied voltage across the network to calculate the similarity coefficients in the LCA equations \cite{Sheridan2016,Sheridan2017}.  Sheridan \mbox{\etal} used a microprocessor to implement the majority of the LCA, and did not include comprehensive throughput and power information \mbox{\cite{Sheridan2016,Sheridan2017}}. However, a fundamental result of their work is that memristive devices have sufficient resolution and accuracy to implement LCAs on real devices \mbox{\cite{Sheridan2016}}.  Our work extends their work by proposing a means of implementing the entirety of the LCA on the same chip as the memristive crossbar with few additional components.  The Sheridan \etal\ work also advocated using a {\em Winner-Take-All} (WTA) approach to training the weight matrix: rather than updating the weights for all columns participating in a reconstruction, they only updated the largest contributor \cite{Sheridan2016}.  While effective, using WTA was motivated largely by the supposition that a single neuron's firing would dominate the response to most stimuli; however, with inhibition and larger, more complicated inputs, this is not the case.

Also noteworthy is recent work by Tang \etal\ which extends the mathematical justification of spiking LCA networks \cite{Tang2017spikingLca}.  Their work is largely theoretical, not corresponding to a single hardware but rather addressing the mathematics of a spiking network used to implement the LCA.  Similar to \cref{sec:model} and Shapero \etal's SLCA \cite{Shapero2013}, they model spikes as rate-based entities, but emphasize proofs of convergence over specific designs.  In contrast, our work makes use of a statistical model to define behavior which is hardware friendly and shown to be empirically useful (\cref{sec:results}).

\section{Model}\label{sec:model}

\fig[label=fig:sslca]
    {\graphic{figs/sslcaOverview.svg}}
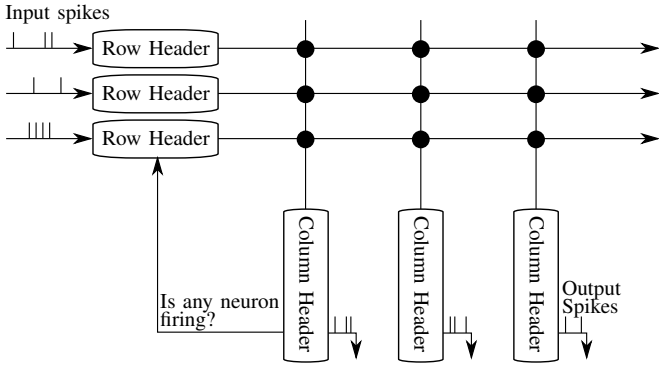
    {High-level architecture for the SSLCA.  During inference, input spikes pass through a Row Header.  Voltage is forwarded from the Row Headers to a nanowire crossbar with memristors at each junction.  Current is allowed to pass through each memristive junction and is used to charge or discharge an LIF neuron in each Column Header (\cref{fig:sslca-i}).  When any LIF neuron spikes, an output spike is propagated and inhibitory forces are passed back through the crossbar to the Row Headers.  Only a single shared bit (``Is any neuron firing?'') is required.  The count of output spikes across any given time window describes the sparse code for the input pattern seen during that time window.}

\fig[label=fig:sslca-sparsecode]
    {\graphic{figs/spiketest0_signal.pdf}}
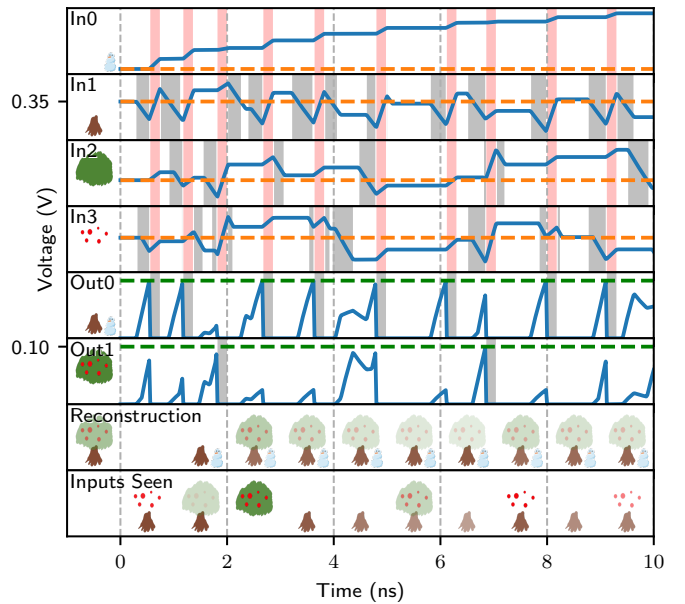
    {Simulated voltage traces within our architecture, using perceptual icons as guides.  Actual model demonstrated is the inhibited SSLCA from \cref{sec:model:sslca-inhib}.  This network has 4 inputs and 2 output neurons; the input vector passed is $[0, 1, 0.5, 0.5]$, the first output neuron responds to the first two inputs, and the second output neurons responds to the last two inputs.  Shaded regions are spikes; red shaded regions demonstrate input regions ignored due to output spikes.  Input voltages shown are the charge on the inhibition capacitors, and the orange dashed line is the inhibition threshold.  Output voltages shown are the charge on the neuron capacitors, and the green dashed line is the firing threshold.  The ``Reconstruction'' row demonstrates the input (far left), and then reconstruction that would be generated if the algorithm stopped at \SI{0.5}{\nano\second}, \SI{1.5}{\nano\second}, etc.  The ``Inputs Seen'' row demonstrates the uninhibited inputs seen between the $i-1$th output spike and the $i$th output spike; that is, it is the total activity that caused an output spike.  Generally, input activity accumulates for inputs that are currently underrepresented in the output, and the output neuron that best represents that difference between input activity and output representation fires, charging inhibition capacitors that prevent that region of the input from being represented again too soon.  More details can be found throughout \cref{sec:model}.}

In light of issues with previous hardware implementations and the potential benefits of sparse coding algorithms discussed in \cref{sec:intro,sec:related}, we set out to develop the simplest architecture for sparse coding that would exhibit \bigOn\ scaling, utilize inhibition, and emphasize low-power operation.  While any device whose resistance can be modified in-situ would suffice, memristors from Lu \etal's group were chosen due to their nanoscale form factor and ability to be fabricated in tight crossbars \cite{Sheridan2016}.  These devices additionally exhibit a low on:off ratio, which has been associated with devices that possess better long-term storage and analog qualities \cite{Sheridan2016,Woods2015}.  It is also worth noting that while the internal state of memristive devices could change at any voltage, it changes very slowly at the \mbox{\SI{0.7}{\volt}} used during inference for the SSLCA.  Further discussion of the time-varying qualities of memristive devices can be found in \mbox{\cite{Woods2015}}; for this architecture, we generally assume that the memristive devices used are static during the inference step due to the low voltages used, and are adjusted in a separate training step that implements Oja's rule, or in a loading process that configures the chip with weights learned during offline training.

As an initial step, we established that the chosen architecture should fit the form shown in \cref{fig:sslca}.  Assuming good accuracy could be derived, such an architecture would be sufficient for implementing sparse coding with the desired traits.  Such an architecture would clearly exhibit \bigOn\ scaling.  Inhibition could be implemented with a backwards-pass through the same crossbar used to charge the output neurons.  Low-power operation would stem from the simplicity of the architecture, its good scaling properties, and an innovation on the way the neurons were integrated into the architecture.

Sparse coding may be realized with this architecture as demonstrated in \cref{fig:sslca-sparsecode}.  Input spikes were chosen due to biological inspiration, the promise of lower power consumption, and also partially because memristors exhibit vastly different resistances at different voltages; using spikes rather than voltage-scaled inputs helps to avoid this situation \cite{Woods2015}.  Upon reaching the Row Headers, each input spike (gray-shaded regions) would be converted to a voltage and would charge the state of any output neurons who encode activity from that input.  As with most spiking algorithms, our proposed network would deal with the statistics of populations of spikes rather than the interactions of individual spikes.  Thus, input spikes might be produced and converted in any fashion so long as the expected voltage of the input line scales linearly with the activity of the input.  The amount of charge received by each output neuron would be controlled by a memristive device at nanowire junctions between the Row and Column Headers.  These memristive devices form a pattern of conductances for each column, constituting its {\em Receptive Field} (RF).  Patterns of input spikes will produce more charge in output neurons whose RFs align with the input pattern.  When an output neuron is charged beyond a threshold by a sufficient quantity of input spikes, that output would produce an output spike (gray-shaded region), representative of the inputs seen so far.  As those inputs are then represented in the pattern of output spikes, all output states would reset when the output spike occurred.  To achieve sparse coding, while the output spike was active, energy would flow backwards through the crossbar, charging inhibition capacitors that inhibit well-represented input spikes from propagating until more spikes have been seen than represented in the output.  As a result, a representative encoding of the input would be produced by counting output spikes, and this encoding would be sparse as a function of both the limited number of output spikes collected and the fact that each output represents a population of input elements.  This figure is referred to throughout this section to explain details of the SSLCA's implementation.

Neurons in this architecture differ from previously proposed architectures.  Like prior work, the Column Headers implement {\em Leaky-Integrate-and-Fire} (LIF) neurons \cite{Payvand2015,Shapero2013}.  They consist of a capacitor that charges (integrates) input events as they are active, and discharges (leaks) when input events are inactive.  In contrast to those architectures, which use a separate resistor to control the leak rate, this architecture's LIF neurons both accrue and dissipate charge via the memristive crossbar.  In addition to requiring fewer components, this configuration has proven more tolerant of un-normalized receptive fields, a phenomenon discussed in \cref{sec:model:uninhib}.

The derivation of necessary parameters was broken down into two stages: calculations without inhibition, and an extension of those calculations to incorporate inhibition.  This divide was necessary to ensure the solution was tractable, and had the added benefit of deriving two versions of the architecture which were used to demonstrate the benefits of inhibition.

\subsection{Uninhibited SSLCA}\label{sec:model:uninhib}

To begin the derivation for the Uninhibited SSLCA, we start with the equation for Rozell \etal's LCA, \cref{eq:lca}, and remove the inhibitory term.  What remains is a leaky dot product, with no \bigOnn\ scaling problem.  However, the resulting equation also no longer possesses optimality guarantees about the quality of its sparse code.

Oja's rule, used in this work to train each neuron's RF, can be used to somewhat remedy the missing inhibitory term by adjusting multiple RFs to work together to reconstruct the input without inhibition \cite{Oja1982}.  Briefly, Oja's rule is that a neuron's receptive field will change proportionally to its output activity multiplied by the difference between the original input and the LCA's reconstruction: $\Delta w_{i, j} = \eta a_j r_i$, where $r_i = x_i - \sum_j w_{i, j}a_j$.  This is identical to gradient descent with a loss function of $r_i^2$.  When both $w_{i, j}$ and $y_j$ are finite and bounded, as in a spiking network using memristors, this equation naturally reduces the weight for inputs that were over-represented, and increases the weight for inputs that were under-represented.  Furthermore, increasing a weight results in an increase of the corresponding spike count.  As a result, any given input produces a natural equilibrium of weight and spike count values.  Either might saturate, but that is not a problem aside from the loss of some of the input's magnitude.

Even with this compensation, using a leaky dot product for sparse coding would be difficult in hardware; consider the angle property of the dot product between two arbitrary vectors $\mathbf{X}$ and $\mathbf{Y}$:

\begin{align}
    \mathbf{X} \cdot \mathbf{Y} &= |\mathbf{X}||\mathbf{Y}|cos(\theta). \label{eq:model:dot}
\end{align}

From \cref{eq:model:dot}, a larger magnitude in either vector could be used to compensate for a larger difference in angle.  In other words, a maximally-conductive RF would generate more current than an RF that is a better match.  Prior works have solved this issue by normalizing each RF \cite{Sheridan2016}.  Instead, we decided to solve this problem by creating a negative stimulus via inactive input channels, achieved by grounding them rather than using high impedance.  The current through these channels would be proportional to the RF, meaning that missing activity where the RF is conductive would lead to a higher penalty.  As a result, a maximally conductive RF is perfectly fine within the system.  It will respond to inputs that are maximally valued, and not respond to inputs with a shape better matched by a different RF.  Coupled with Oja's rule, a maximally conductive RF will adapt to not be maximally conductive if it could better represent a wider range of inputs with a more specialized shape.  This is the reason that capacitors in this network are placed directly on the crossbar, rather than behind a diode or equivalent.  Placing them directly on the crossbar allows accrued charge to dissipate when an input row is inactive.  This phenomenon can be witnessed in the output voltage traces in \cref{fig:sslca-sparsecode} at around \SI{4.5}{\nano\second}.

Using the layout from \cref{fig:sslca}, the Row Headers for the uninhibited SSLCA are simple passthroughs (input spikes are directly connected to the crossbar), and the Column Headers are simply a capacitor and a Schmitt trigger that drains all capacitors once any one neuron's voltage exceeds $V_{fire}$ volts.  The partial derivative of any neuron's voltage is therefore:

\begin{align}
    C\frac{\partial V_{neuron}}{\partial t} &= \sum_i{ (V_i(t) - V_{neuron})G_i }, \label{eq:model:sslca-base}
\end{align}

\noindent where $C$ is the capacitance of the capacitor, $V_{neuron}$ is the current voltage of that capacitor, $V_i(t)$ is the $i$th input's voltage at time $t$ (one of $V_{cc}$ or \SI{0}{\volt} depending on whether it is currently spiking or not), and $G_i$ is the conductance of the memristive device connecting the nanowires of the $i$th Row Header and the neuron in question's Column Header.

This equation can be better reasoned about by assuming an input row's voltage $V_i$ spikes to voltage $V_{cc}$ with a mean activity of $K_i$, meaning that at any point in time the voltage is $V_{cc}$ with probability $K_i$ and grounded with probability $1 - K_i$.  For non-spiking inputs, such as when converting a stored image to a format consumable by the SSLCA, the desired duty cycle $K_i$ may be produced through a simple multiplication.  First, establish a desired maximum input spike duty cycle, $K_{max}$, and then multiply this max cycle by the analog input (assumed to be bounded on $[0, 1]$) to produce $K_i$, the duty cycle of the input's spiking representation.  Any method of generating spikes with the given duty cycle might be used; what is important is the expected value of the input line voltage.  In \cref{fig:sslca-sparsecode}, relative values of $K_i$ can be seen as the portion of each input's area covered by spike activity (the gray-shaded regions).

Using an expectation to replace $V_i(t)$ with $E[V_i(t)] = K_iV_{cc}$, \mbox{\Cref{eq:model:sslca-base}} can then be reduced via the Laplace transform to:

\begin{align}
    Q_1 &= \sum_i{G_i}, \nonumber\\
    Q_2 &= V_{cc}\sum_i{K_iG_i}, \nonumber\\
    C\frac{\partial V_{neuron}}{\partial t}
            &= Q_2 - Q_1V_{neuron}, \nonumber\\
    V_{neuron}(t) &= \frac{Q_2}{Q_1}(1 - e^\frac{-tQ_1}{C}) + V_{neuron, t=0} e^\frac{-tQ_1}{C}, \label{eq:model:sslca-charge}
\end{align}

\noindent where $V_{neuron, t=0}$ is the neuron's voltage at $t = 0$.  $Q_1$, the column's total conductance, and $Q_2$, a matching metric between the stored RF and the input pattern, arise as intuitive factors that affect the neuron's state.  To establish the necessary values for $C$ and $V_{fire}$, $Q_1$ and $Q_2$ need to be derived in a way that produces good results for the network's ``average case.''  Empirically, we found that assuming both the input and stored RF have binary elements (even for analog problems) produced the best results: the $K$ values are either $1$ or $0$, while the $G$ values are either the minimum or maximum conductance of our memristive devices.

The resulting calculation for $Q_1$ and $Q_2$, required to determine both the network's trigger voltage $V_{fire}$ and neuron capacitance $C$, is described in \cref{alg:model:sslca-q}.  Though these calculations are based on a single sample of $Q_1$ and $Q_2$, our results showed that the network still worked well outside of these ``average cases'' (\cref{sec:results}).

Sparse coding being the goal of this architecture, we also make the assumption that any spike event will reset all neuron charges to \SI{0}{\volt}, implying that each output spike only encodes input activity seen since the end of the previous output spike.  Any input spikes in this reset window will be ignored by the system (shown as red-shaded regions in \cref{fig:sslca-sparsecode}).  Since the SSLCA is designed to produce multiple output spikes across time for a single combination of inputs, this is not a problem as the statistics of the input spikes give equal likelihood of a spike during the reset as during any other period of operation.  Real-time (non-episodic) uses of the SSLCA would also not suffer from this implementation detail, assuming all changes to the input stimulus happen at a significantly lower frequency than the spiking frequency of the system.  This phenomenon could be compared to standard sampling theory: one spike is a sample of the input leading up to that spike, and changing frequencies in the input that exceed some fraction of the frequency of the sample rate cannot be deduced.

The downside to this assumption is that the architecture becomes a one-hot system: a pattern of simultaneously-firing output spikes becomes impossible.  Superficially, this is in contrast to some other work on stochastic computation with spiking neurons \cite{Hamilton2014, Habenschuss2013, Maass2015}.  The network still encodes stochastic information in a single output spike: the input pattern represented is stochastic due to the phase of input spikes' duty cycles, and as such the corresponding output is stochastically selected.  By not allowing a pattern of simultaneous output spikes, the number of representable input patterns in a single event is reduced.  However, due to this stochasticity, we have found that collecting multiple output spikes over a period of time results in a stochastic pattern of output activity that accurately represents the input.  This is functionally identical to the trade-off of memory for time in computation: we are reducing the memory of the momentary output of our architecture in exchange for longer runtime.  For sparse coding, where the resulting code often needs to be stored or otherwise buffered, this is not an issue.  Practically, as in \cref{fig:sslca-sparsecode}, the reconstruction gets progressively better the longer the algorithm runs, with diminishing returns.

With the above assumption, all $V_{neuron, t=0} = 0$, and \cref{eq:model:sslca-charge} can be rearranged to calculate $C$ based on some $Q_1, Q_2, V_{fire}, \text{ and } t_{fire}$, where $V_{fire}$ and $t_{fire}$ are the desired voltage and time at which an output spiking event should occur given the input and stored RF parameters that produce $Q_1$ and $Q_2$:

\begin{align}
    C &= \frac{-t_{fire}Q_1}{\ln\left(1 - V_{fire}\frac{Q_1}{Q_2}\right)}. \label{eq:model:sslca-uninhib}
\end{align}

As $t_{fire}$ can be calculated from the desired hardware clock rate and number of spikes per patch, the remaining parameters needed to fully specify the uninhibited SSLCA are $V_{fire}$, $Q_1$, and $Q_2$.  Our experiments produced the lowest reconstruction RMSEs when $V_{fire}$ is calculated based on a thresholded max voltage from \cref{eq:model:sslca-charge} with a $Q_1$ and $Q_2$ calculated for the desired minimum RF that the resulting sparse code can represent, and when the $Q_1$ and $Q_2$ for the calculation of $C$ come from an average case of the data set used with the network.  The exact procedure followed to calculate these values is described in \cref{alg:model:sslca}.  The algorithm requires knowledge of the expected average value of a stored receptive field, $Rf_{avg}$, as well as an idea of the minimum input intensity that should trigger an output spike, $Rf_{least}$.  After scanning across many different combinations of $Rf_{avg}$ and $Rf_{least}$, we discovered that setting $Rf_{least} = (1 - e^{-1})Rf_{avg}$ typically yielded optimal results with regards to classification accuracy, as can be seen in \mbox{\cref{fig:model:sslca-ratio}}; this relation was used throughout this work.  Though \mbox{\cref{fig:model:sslca-ratio}} displays results only on CIFAR, we found a similar result shape with MNIST when trying different values of $Rf_{avg}$ and $Rf_{least}$.  In the context of \cref{sec:model:sslca-inhib}, the inhibited network, the shape was also similar.

\fig[label=fig:model:sslca-ratio]
    {\graphic{plots/cifar-8x8inhib-score.pdf}}
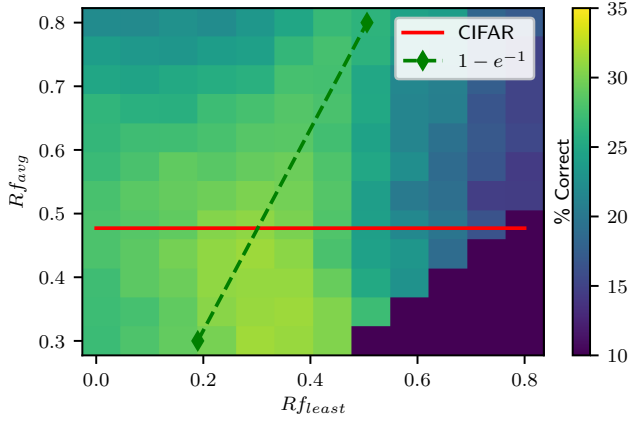
    {Accuracy for the SSLCA network on CIFAR-10 across different values of $Rf_{least}$ and $Rf_{avg}$.  Generally, setting $Rf_{least} = (1 - e^{-1})Rf_{avg}$ was found to be a safe choice.}

\alg[label=alg:model:sslca-q]
    {Process used to determine $Q_1$ and $Q_2$ given stored RF of average, relative conductance $Rf_{stored}$ and a matching input of average, relative intensity $Rf_{input}$.}
    {
        \Input{
            $Rf_{stored}$, the average, relative conductance of the stored RF.  This value must be on the interval $(\frac{G_{min}}{G_{max}}, 1]$; \par
            $Rf_{input}$, the average, relative intensity of all inputs; \par
            $G_{min}$, the minimum conductance of a crossbar device; \par
            $G_{max}$, the maximum conductance of a crossbar device; \par
            $K_{max}$, the proportion of time spent at $V_{cc}$ for an input signal spiking at its maximum rate; \par
            $N$, the number of inputs to the network.
        }
        \Output{$Q_1, Q_2$}
        \Begin{
            \tcp{Assumes that the stored RF consists entirely of elements at $G_{max}$ or $G_{min}$, and that the input pattern matches, but with a scaled intensity of $\frac{Rf_{input}}{Rf_{stored}}$.  This simplification helps the network perform well with high-contrast RFs.}
            $g_{min} \gets \frac{G_{min}}{G_{max}}$; \\
            $I_h \gets \frac{Rf_{stored} - g_{min}}{1 - g_{min}}$; \tcp*[f]{\parbox[t]{1.25in}{\raggedright Portion of inputs at max-intensity.}} \\
            $I_l \gets 1 - I_h$; \\
            $Q_1 \gets NG_{max}Rf_{stored}$; \\
            $Q_2 \gets NV_{cc}G_{max}K_{max}\frac{Rf_{input}}{Rf_{stored}}\left( I_h + I_lg_{min}^2 \right)$. \\
            \tcp*[f]{\parbox[t]{2in}{\raggedright Note that the $g_{min}^2$ term comes from one $g_{min}$ multiplication for $G_{max}$ and another for $K_{max}$.  This is the ``match'' term between the RF and the input.}}
        }
    }
\alg[label=alg:model:sslca]
    {Recommended process for selecting $V_{fire}$, $Q_1$, and $Q_2$, required for the calculation of $C$ from \cref{eq:model:sslca-uninhib}.}
    {
        \Input{
            $Rf_{avg}$, the desired average, relative conductance of a stored RF.  This value must be on the interval $(\frac{G_{min}}{G_{max}}, 1]$; \par
            $Rf_{least}$, the smallest average, relative input intensity is expected to trigger an output spike; \par
            $G_{min}$, the minimum conductance of a crossbar device; \par
            $G_{max}$, the maximum conductance of a crossbar device; \par
            $K_{max}$, the proportion of time spent at $V_{cc}$ for an input signal spiking at its maximum rate; \par
            $N$, the number of inputs to the network.
        }
        \Output{$Q_1, Q_2, V_{fire}$}

        \Begin{
            $V_{fire} \gets (1 - e^{-1})\frac{Q_2}{Q_1}$, with $Q_1, Q_2$ from \cref{alg:model:sslca-q} applied to $Rf_{stored} = Rf_{avg}$, $Rf_{input} = Rf_{least}$, other parameters matching; \\
            $Q_1, Q_2 \gets Q_1, Q_2$ from \cref{alg:model:sslca-q} applied to $Rf_{stored} = Rf_{avg}$, $Rf_{input} = Rf_{avg}$.
        }
    }

Following \cref{alg:model:sslca}, and substituting the resulting values into \cref{eq:model:sslca-uninhib}, all parameters for constructing the uninhibited network are defined, and the network might be built.  Applying voltage spikes to the input lines of magnitude $V_{cc}$ with a maximum duty cycle of $K_{max}$ will cause the best-matching column to spike for $t_{spike}$ seconds; collecting these spikes across a window of time (e.g. $10(t_{fire} + t_{spike})$ for an average of $10$ spikes) will produce a reasonable reconstruction of the input based on the network's receptive fields.  Results with the uninhibited SSLCA can be found in \cref{sec:results}.

\subsection{Adding Inhibition to the SSLCA}\label{sec:model:sslca-inhib}

One of the original requirements deduced at the beginning of \cref{sec:model} was the need for inhibition.  Prior works have shown the need for inhibition in an effective sparse coding system \cite{Woods2015tnano,Querlioz2012}, and \cref{sec:results:cifar:soa} of this work demonstrates this as well.  While works such as that of Shapero \etal\ implemented inhibition by using additional hardware between each pair of neurons \cite{Shapero2013}, leading to \bigOnn\ scaling, the SSLCA is designed in a way that allows for \bigOn\ scaling.

Instead of additional hardware, a percentage of the SSLCA's running time is dedicated to calculating inhibitory forces.  Whenever an output spike is generated, the duration of the output spike is used to pass current from the corresponding column back through the SSLCA's crossbar, charging capacitors in the Row Headers.  Intuitively, the charges on these capacitors indicate how well represented the corresponding input signal is in the current reconstruction; overrepresented input signals will be suppressed.  This is implemented through the Row and Column Headers shown in \cref{fig:sslca-i}.  The effect of this can be seen on the voltage traces of the inhibition capacitors of \cref{fig:sslca-sparsecode} during the red-shaded regions: inputs corresponding to the currently-firing output charge their inhibition capacitors more quickly than inputs that are poorly-represented by the spiking output.

\fig[label=fig:sslca-i]
    {\graphic{figs/sslcaInhib.svg}}
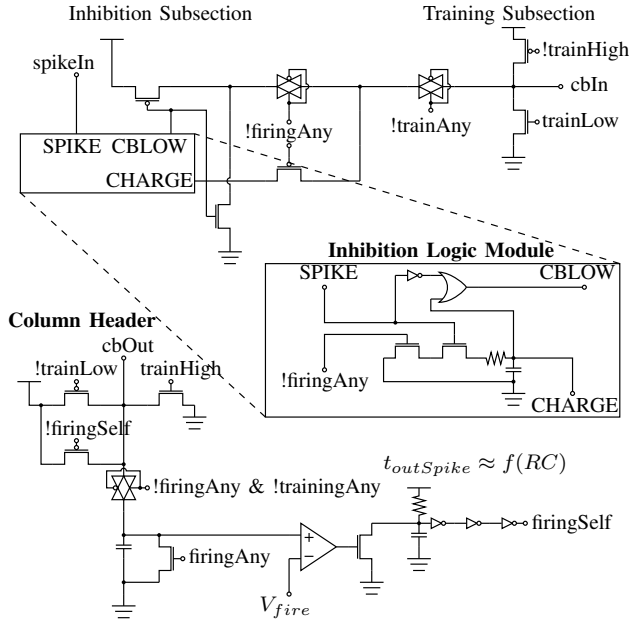
    {The Row and Column headers needed to add inhibition to the SSLCA.  The Row Header's responsibilities are to stop input spikes from reaching the crossbar when they are inhibited, and to keep track of the current state of the inhibitory forces.  An Inhibition Logic Module is diagrammed as broken out from the main circuit for space reasons.  The CHARGE port is responsible for sinking current from the crossbar when an output spike has occurred, and in turn charges the capacitor in the Inhibition Logic Module which prevents subsequent spikes from applying a voltage on the crossbar.  After enough input spikes occur, the capacitor becomes sufficiently drained to apply voltage to the crossbar once more.  The Column Header is much simpler and uses a transmission gate to direct current to and from the neuron's state capacitor.  When any neuron fires, the capacitor is drained, and in the same column, $V_{cc}$ is applied to the crossbar.  A simple RC circuit cleaned up by several NOT gates is responsible for the output spike.}

The Column Header for the Inhibited SSLCA is almost identical to that of the Uninhibited SSLCA: a standard LIF neuron setup, with the state capacitor connected directly (through a transmission gate) to the nanowire crossbar rather than being buffered.  A crude schmitt trigger setup ensures that all output capacitors drain sufficiently when any neuron fires (firingAny), resetting the spike potentials.  Additionally, if the current column is firing (fireSelf), the crossbar column is pulled up through a transistor.  As a result, during each output spike, an inhibition current will flow back through the memristive crossbar into the Row Headers.  The current flowing into each Row Header will be proportional to the receptive field of the firing column.

The Row Header is more complicated, but the important aspect is that a capacitor storing the inhibition state discharges whenever an input spike arrives, and charges whenever an output spike occurs.  In this way, the inhibition capacitor contrasts how much the input is represented in the output (increasing inhibition voltage) with the actual activity of the input (decreasing inhibition voltage).  The capacitor is charged through the crossbar junctions; the resistor for discharging the capacitor, referred to as $R_{inhib}$, is the one in the labeled Inhibition Logic Module.  The stored inhibition state, when above $\frac{V_{cc}}{2}$, prevents input spikes from reaching the crossbar.  $\frac{V_{cc}}{2}$ is chosen as it maximizes the linearity of the inhibitory response, since both charging and discharging occur at the same point on the exponential function (\cref{eq:model:inhib-first}).

Calibrating this architecture requires specifying both the capacitor, $C_{inhib}$, and the resistor, $R_{inhib}$, in the Inhibition Logic Module (\cref{fig:sslca-i}).  Ideally, a sparse coding algorithm should produce a stable, one-hot response to an input that exactly matches any of the stored RFs, and should combine several outputs when representing inputs that do not match a stored RF exactly.  For simplicity, we focused on tuning the inhibitory components of the network to an input that matches the stored conductance of an RF, similarly to \cref{alg:model:sslca-q,alg:model:sslca}.  Additionally, we make room for inhibition in the spike cycle by using a neuron capacitance of $C_{cb} = f(C)$, where $C$ is from \cref{eq:model:sslca-uninhib} and $f(C)$ is an arbitrary function with value $0 < f(C) < C$.  Using $R_{cb}$ as the equivalent resistance of the memristive device used to charge the inhibitory force, and $R_{inhib}$ as the resistance in the Inhibition Logic Module, we can write a few equations to describe the inhibition voltage $V_{i}$ for a specific input $i$ both before an output spiking event ($V_{i,pre}$) and after an output spiking event ($V_{i,post}$):

\begin{align}
    A &= \frac{1}{R_{cb}C_{inhib}}, \nonumber\\
    B &= \frac{K_i}{R_{inhib}C_{inhib}}, \nonumber\\
    V_{i,pre} &= V_{i,0}e^{-t_{fire}B}, \nonumber\\
    V_{i,post} &= V_{cc} + (V_{i,pre} - V_{cc})e^{-t_{spike}A}, \label{eq:model:inhib-first}
\end{align}

\noindent where $t_{spike}$ is the duration of an output spike, $K_i$ is the portion of the time that the input being tracked is active, and $V_{i,0}$ is the voltage after an output spike.  For a stable system with a uniform firing rate, $V_{i,0} = V_{i,post}$, and we are left with:

\begin{align}
    V_{i,0} &= V_{cc} + \left(V_{i,0}e^{-t_{fire}B} - V_{cc}\right)e^{-t_{spike}A} \nonumber\\
            &= \frac{V_{cc}\left(1 - e^{-t_{spike}A}\right)}{1 - e^{-t_{fire}B - t_{spike}A}}. \label{eq:model:inhib-partial}
\end{align}

With the inhibition voltage after a spike defined, one issue remains: so long as $V_{i,0} > \frac{V_{cc}}{2}$, the desired $t_{fire}$ will no longer match $t_{fire}$ without inhibition.  There is always a period of time during which input spikes are inhibited, inflating $t_{fire}$.  We label this period of time as $t_{inhib}$, and rewrite $t_{fire}$ as $t_{inhib} + t_{collect}$, allowing \cref{eq:model:inhib-partial} to be rewritten and a second equation for $V_{i,0}$ to be written by integrating backwards to $V_{i,0}$ from $\frac{V_{cc}}{2}$.  These two equations are then combined to make a single equality, the solution of which indicates adequate values for $R_{inhib}$ and $C_{inhib}$:

\begin{align}
    V_{i,0} &= \frac{V_{cc}\left(1 - e^{-t_{spike}A}\right)}{1 - e^{-t_{collect}B - t_{spike}A}}, \nonumber\\
    V_{i,0} &= \frac{V_{cc}}{2}e^{t_{inhib}B}, \nonumber\\
    \frac{V_{cc}}{2}e^{t_{inhib}B} &= \frac{V_{cc}\left(1 - e^{-t_{spike}A}\right)}{1 - e^{-t_{collect}B - t_{spike}A}}. \label{eq:model:sslca-rinhib}
\end{align}

Unfortunately, this formulation leaves two new variables, $t_{inhib}$ and $t_{collect}$.  Additionally, were we to use the original capacitance calculated in \cref{eq:model:sslca-uninhib}, we would miss the desired $t_{fire}$ due to the added time for inhibition.  To solve all of these problems, we set $C_{cb} = f(C) = \frac{C}{2}$.  $t_{collect}$ is then solved for using the new neuron capacitance $C_{cb}$ and $Q_1, Q_2$ from \cref{alg:model:sslca-q} using $Rf_{stored} = Rf_{input} = Rf_{avg}$.  $t_{inhib}$ is then solved by subtracting $t_{collect}$ from $t_{fire}$.  The remaining variable, $R_{inhib}$, is solved for by taking the log of both sides of the above equality (\cref{eq:model:sslca-rinhib}), squaring the result, and minimizing the resulting function via Python's {\em scipy.optimize.minimize}, ensuring a near-zero result \cite{scipy}.

Examples of the results from \cref{alg:model:sslca} plus the inhibition transformations ($C_{cb} = f(C)$) can be seen in \cref{tab:model:sslca-values}.  Notably, $N = 192$ corresponds to an input image of dimension $8\times 8\times 3$, while $N = 48$ corresponds to an input dimension of $4\times 4\times 3$.  Row 1 is similar to the settings used in most of our experiments.  While $C_{cb}= 1200$\SI{}{\femto\farad} is significant, this number could be greatly reduced by future memristive technologies with greater resistance (rows 2 and 3).  Rows 4 and 5 highlight that a higher ratio of $G_{max}$ to $G_{min}$ results in a higher $V_{fire}$, which would be helpful to overcome the comparator's input offset voltage and would allow the algorithm to better represent zero weights in each neuron's RF; both of these would increase the algorithm's effectiveness.  A lower $G_{max}$ may be artificially imposed on the network if the circuit designer wants less capacitance and is willing to sacrifice some of the accuracy that comes from a high $G_{max}$ to $G_{min}$ ratio.  Rows 6 through 8 demonstrate the effects of fewer inputs, and of varying $Rf_{avg}$, the expected average stored RF in the network.  As written, $Rf_{avg}$ is also treated as the average input to the network; for very low $G_{max}$ to $G_{min}$ ratios, this might not make sense, and the actual average input value should be added as a separate input to \cref{alg:model:sslca} and used in the final calculation of $Q_1$ and $Q_2$ to correct for the difference.

\wtable[label=tab:model:sslca-values, padh=0.25em]
    {Example Parameters for Inhibited Network}
    {|r
     |S[table-format=3.0,   fixed-exponent=0, table-omit-exponent=true, round-mode=figures, round-precision=3, zero-decimal-to-integer=true]
     |S[table-format=0.2, fixed-exponent=0, table-omit-exponent=true, round-mode=figures, round-precision=2]
     |S[table-format=1.3, fixed-exponent=-6, table-omit-exponent=true, round-mode=figures, round-precision=2]
     |S[table-format=2.2, fixed-exponent=-6, table-omit-exponent=true, round-mode=figures, round-precision=2]
     |S[table-format=3.0, fixed-exponent=-3, table-omit-exponent=true, round-mode=figures, round-precision=2]
     |S[table-format=4.0, fixed-exponent=-15, table-omit-exponent=true, round-mode=figures, round-precision=2]
     |}
    {%
        {Row} & {$N$} & {$Rf_{avg}$} & {$G_{min}$ (\SI{}{\micro\siemens})} & {$G_{max}$ (\SI{}{\micro\siemens})} & {$V_{fire}$ (\SI{}{\milli\volt})} & {$C_{cb}$ (\SI{}{\femto\farad})} \\
        \hline

        1 &  192.0 &    0.4 &  4.830918e-06 &  1.923077e-05 &  0.087154 &  1.158099e-12 \\
        2 &  192.0 &    0.4 &  4.830918e-07 &  1.923077e-06 &  0.087154 &  1.158099e-13 \\
        3 &  192.0 &    0.4 &  4.830918e-08 &  1.923077e-07 &  0.087154 &  1.158099e-14 \\
        4 &  192.0 &    0.4 &  4.830918e-08 &  1.923077e-06 &  0.134582 &  1.158099e-13 \\
        5 &  192.0 &    0.4 &  4.830918e-08 &  1.923077e-05 &  0.139325 &  1.158099e-12 \\
        6 &   48.0 &    0.4 &  4.830918e-06 &  1.923077e-05 &  0.087154 &  2.895247e-13 \\
        7 &   48.0 &    0.6 &  4.830918e-06 &  1.923077e-05 &  0.116431 &  4.342871e-13 \\
        8 &   48.0 &    0.8 &  4.830918e-06 &  1.923077e-05 &  0.131069 &  5.790494e-13 \\
        }

To validate this network design, we investigated different parameters other than $Rf_{avg}$ for $Rf_{stored}$ and $Rf_{input}$ when optimizing the inhibitory response.  \Cref{fig:model:sslca-rinhib} demonstrates the results of this: while different combinations require different values of $R_{inhib}$ to be completely accurate, choosing a single, median value works well in practice.

\fig[label=fig:model:sslca-rinhib]
        {\graphic{plots/r-inhib-err.pdf}}
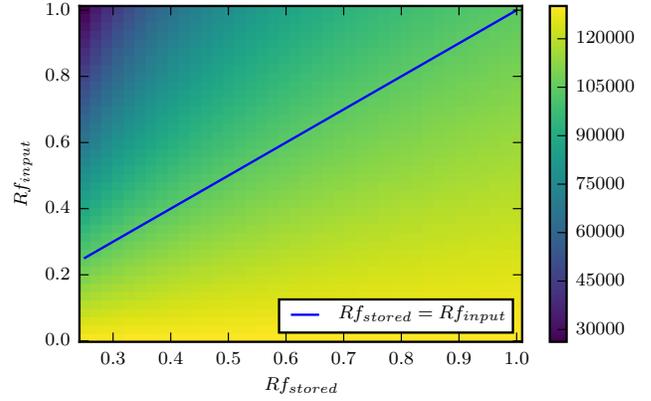
        {Values of $R_{inhib}$ needed to achieve the desired spike rate with different $Rf_{stored}$ and $Rf_{input}$ values (the spike rate scales with the ratio of $Rf_{input}$ over $Rf_{stored}$).  Ideally, the resulting plot would be flat, indicating a single value of $R_{inhib}$ is sufficient for all cases.  Since it is not flat, areas with larger than the chosen $R_{inhib}$ will spike slower than expected, and areas with a smaller value will spike faster.  In practice, we use $Rf_{avg}$ for both (the blue line); receptive fields with a high stored value and a low input will under-spike, which should not be an issue as those regions should be better-covered by another neuron.}



\subsection{Training}\label{sec:model:training}

The networks we used were trained using the ADADELTA algorithm in tandem with Oja's rule across two epochs of the training data, an identical approach to our prior work \cite{Zeiler2012,Oja1982,Woods2015tnano}.  When considering the reconstructions for Oja's rule, we used the ratio of the conductance of each memristive device to $G_{max}$, the maximum expected conductance of a crossbar device.  This approach limited the minimum representation of each input element in an RF to the inverse of the conductance on/off ratio of the memristive device.  For our experiments, we used the Yang \etal\ device which featured a conductive on/off ratio of around $4$ at \SI{0.7}{\volt} \cite{Woods2015}.  We also tried training without this limitation (allowing the learned weight to drop all the way to $0$, even though the device conductance would be set to $0.25$), but did not find such a change to impact accuracy, although it did affect the RMSE between the input and the reconstruction.  Since the logical minimum does not affect the programmed conductance, this makes sense: the resulting sparse code is unchanged.  The benefit of training with a non-zero minimum representable value is that the training could be done using only the memristive crossbar, without supplemental memory.

Homeostasis was used during training to encourage the network to use all available neurons, similar to prior work by Querlioz \etal\ \cite{Querlioz2013}.  If a neuron had not produced an output spike after several patches, $V_{fire}$ was lowered for that neuron to encourage it to spike.  This behavior was disabled for evaluating accuracy and RMSE.

For this work, all conductances were represented as analog values.  We have conducted prior research that assumed a lower resolution of conductances would be achievable \cite{Woods2015tnano}.  Currently available literature has shown that memristors might be trained within 1\% of a target resistance \cite{Alibart2012}, which is much better than the 4-bit resolution needed for good performance with neuromorphic algorithms \cite{Woods2015tnano, Pfeil2012}.  Note that a 4-bit resolution corresponds to $\pm 3.1\%$ write accuracy.

\subsection{Models Used for Power and Accuracy Comparisons}\label{sec:model:acc}

The SSLCA was simulated algorithmically based on the above equations and algorithms.  The simulator was written as a hybrid event/time-based simulator based on the maximum of the next predicted spiking event and a small window of time (\SI{2e-18}{\second}).  Traces from this simulator are shown in \cref{fig:sslca-sparsecode}.  An identical setup was used to produce that figure as the standard setup for the MNIST and CIFAR experiments, with the exception that \cref{fig:sslca-sparsecode} only used a network with 4 inputs and 2 outputs.  Even so, the function of the network is identical for larger networks.  Unless otherwise specified, our algorithm was configured to collect an average of 10 spikes per exposed image, based on \cref{fig:model:compare}.  Accuracy was computed with a {\em Single-Layer Perceptron} (SLP) network that was trained to associate resulting sparse codes with the category that generated them.  This setup is efficient to compute, but does not rival the accuracy of a state-of-the-art deep learning architecture.  A deep learning classifier was investigated in \cref{sec:results:cifar:soa}.

While crossbar and capacitor power were calculated through these simulations, comparator power for the column headers was derived by simulating the \SI{5}{\giga\hertz} comparator from Xu \etal\ 2011 at \SI{4}{\giga\hertz} using a \SI{0.7}{\V} power supply, implemented with \SI{45}{\nano\meter} CMOS transistors using the Predictive Technology Model published by Zhao \etal\ in 2006 \cite{Xu2011,Zhao2006}.  It was found that, per column, this setup added \SI{2.2}{\micro\watt}.

Since we used Xu \etal's comparator at \SI{4}{\giga\hertz} \cite{Xu2011}, we configured the networks for an average spike accumulation period ($t_{fire}$) of \SI{0.8}{\nano\second} and an output spike duration ($t_{spike}$) of \SI{0.2}{\nano\second}.  Thus, the 10 average output spikes occur roughly once every $t_{fire} + t_{spike} = $\SI{1}{\nano\second}.  Regardless of the actual number of spikes, the algorithm stops after \SI{10}{\nano\second}, and moves on to the next image.  Input spikes were considered with a firing time of \SI{0.4}{\nano\second} and a maximum active duty cycle of $K_{max} = 0.5$ unless otherwise noted.  Assuming a data set with an average input intensity of $0.5$, which is similar to the average of CIFAR-10, this means that each input spikes, on average, once every $\frac{0.4}{0.5K_{max}} = $\SI{1.6}{\nano\second}.  Coupled with the time represented by each output spike, $t_{fire} = $\SI{0.8}{\nano\second}, we infer that each output spike experienced spikes from only half of the active inputs, on average.  Note that the input spike generation method is not important so long as the expectation of the input voltage is maintained.  Our model used a simple uniform-noise-driven model to produce gaps between spikes such that the expected voltage on the spiking line was proportional to the corresponding image element's intensity.

\subsection{Example Code Availability}
The simulation implementation used in this work was made available on Github at \url{https://github.com/wwoods/tlab_sslca}.

\section{Results}\label{sec:results}

The SSLCA, SLCA, and LCA were tested with two different data sets to demonstrate the relative performance of the SSLCA.  Reported RMSE values were generated as though zero were representable, and accuracies were from an SLP (discussed in \cref{sec:model:training,sec:model:acc}).  Experiments were run either \SI{12}{} times, or until $\pm$\SI{10}{\%} accuracy was achieved with \SI{95}{\%} confidence as per \cite{Driels2004}.

To show that our assumptions and simplifications did not result in significantly worse accuracy than the algorithms from which the SSLCA was derived, all results were compared with both the LCA and Shapero \etal\cite{Shapero2013}'s SLCA.  An accuracy comparison across different numbers of output spikes can be seen in \cref{fig:model:compare}.  The LCA implementation is from equation (3.1) of Rozell \etal's paper \cite{Rozell2008}; the SLCA implementation consisted of equations (5)-(7) in Shapero \etal's paper \cite{Shapero2013}.  Note that only the outputs of the SLCA network are spiking, while its inputs are constant voltages.  Our work dealt with both spiking inputs and outputs.

Power numbers for the LCA come from (13) of Shapero \etal's 2012 work \cite{Shapero2012} and scale as \bigOnsqrtn.  Power numbers for the SLCA come from Shapero \etal's 2013 work \cite{Shapero2013}, and as that work included no built-in Vector Matrix Multiplier (VMM) as our algorithm does, we added the power from a memristor-based VMM to its figures.  The throughputs of each of those architectures were several orders of magnitude lower than the SSLCA's (\cref{fig:related:compare}).

\subsection{CIFAR-10}
The first dataset, CIFAR-10, consisted of \SI{60000}{} $32\times 32$ RGB images, each containing one of \SI{10}{} classes of objects \cite{Krizhevsky2009}.  For faster simulation and to demonstrate the scalability of each algorithm, these were scaled down to both $3\times 3$ and $8\times 8$.  As the CIFAR-10 dataset contains equal numbers of each class, a simple accuracy was used to evaluate each algorithm's abilities.

\fig[label=fig:model:compare]
    {\graphic{plots/spike-cifar-combine.pdf}
        \graphic{plots/spike-cifar-power.pdf}}
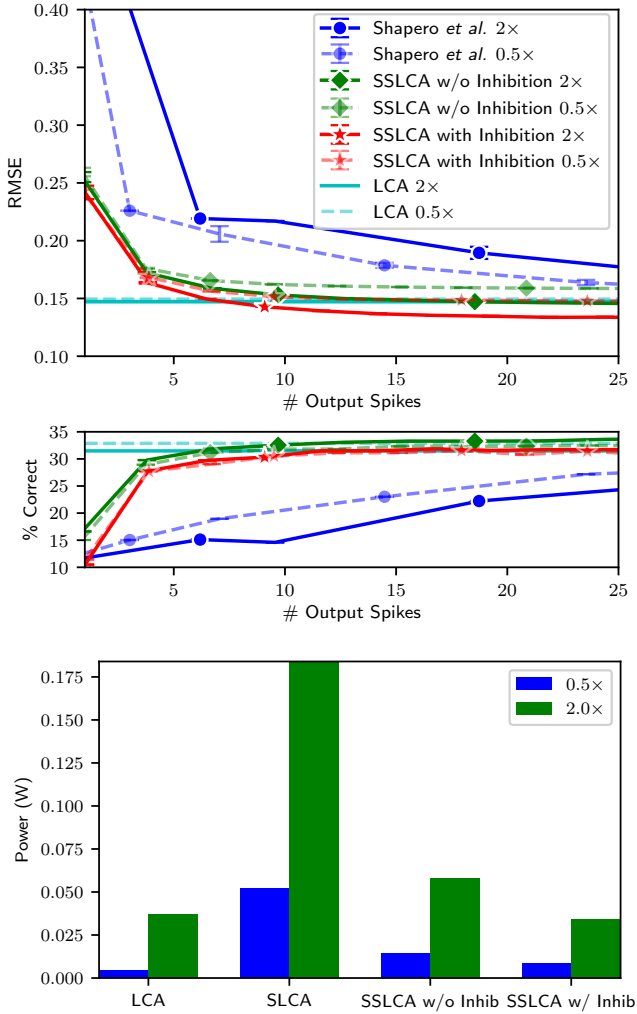
    {Comparison of the LCA \cite{Rozell2008}, SLCA \cite{Shapero2013}, and the SSLCA on CIFAR-10 scaled to $8\times 8$; suffixes indicate completeness ($2\times$ indicates $384$ neurons, while $0.5\times$ indicates $96$ neurons).  While  the SLCA achieves lower RMSE with significantly more spikes (around \SI{100}{}), for practical numbers of spikes the SSLCA produced much better results.  The LCA performed better classification with fewer output neurons because it had slightly less output activity, which with a shallow classifier is more effective.  A lower RMSE is more important for deep learning, seen in \cref{fig:model:cifar:soa}.  While the LCA displayed promising power statistics for this problem, its throughput was four orders of magnitude smaller.}

\subsubsection{Accuracy}
\fig[label=fig:model:cifar:score]
    {\graphic{plots/cifar-combine-score8x8.pdf}}
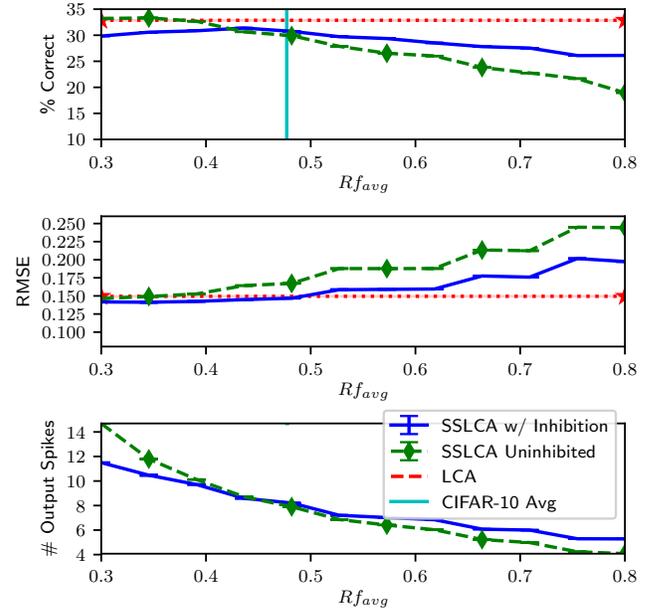
    {SSLCA accuracy targeting 10 spikes on CIFAR-10 rescaled to $8\times 8$ with and without inhibition, compared to LCA.  Lower $Rf_{avg}$ tended to produce lower RMSE due to increased activity in the resulting sparse code, and increased spike count (since the input intensity is greater than the target, spikes happen more frequently than calibrated).  Inhibition ubiquitously reduced the RMSE, although with an SLP, its classification accuracy was less than the uninhibited version for darker receptive fields.  See \cref{sec:results:cifar:soa} for the impact of RMSE when using a deep classifier.}

Compared to an optimal, analog implementation of the LCA, the SSLCA with inhibition matched performance on the $8\times 8$ rescale of CIFAR-10 with a \SI{3}{\percent} relative loss in accuracy (\SI{33}{\percent} vs \SI{32}{\percent}; \cref{fig:model:cifar:score}).  The uninhibited SSLCA always produced a worse reconstruction than its inhibited counterpart, although for low $Rf_{avg}$ (and correspondingly a higher number of spikes per patch) its classification accuracy was better with the simple SLP classifier.  The trained network had an average spike count of \SI{8}{} even though the architecture was configured for \SI{10}{} spikes.  The spike duty cycles were $K_{in} = 0.5$ and $K_{out} = 0.2$, where $K_{in}$ is the maximum duty cycle of input spikes (and will be scaled by each input's intensity), and $K_{out}$ is the expected duty cycle of the output spikes.  That is, an output spike spikes for $K_{out}(t_{fire} + t_{spike})$.

\fig[label=fig:model:cifar-3v8]
    {\graphic{plots/cifar-combine-score3v8.pdf}}
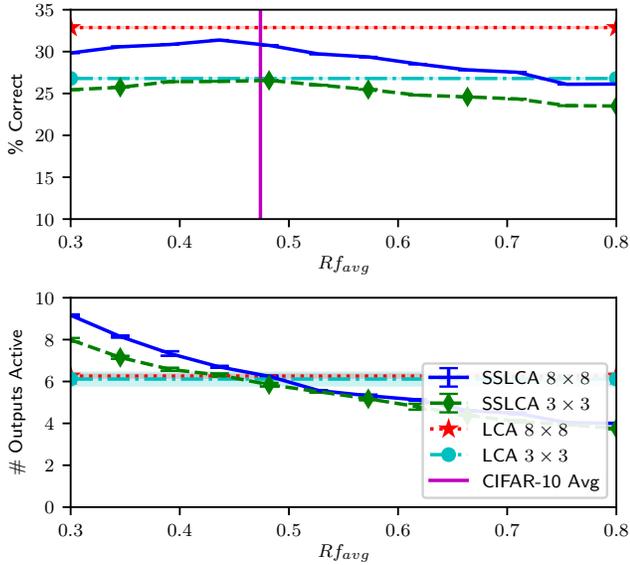
    {A look at the difference between CIFAR-10 scaled to $3\times 3$ and $8\times 8$.  In both cases, the SSLCA approached the LCA's accuracy.  Unfortunately, the required setting of $Rf_{avg}$ to maximize accuracy is not intuitive.  The lower plot shows the sparsity of the output for each configuration; like the original LCA's $\lambda$ threshold, a combination of the $Rf_{avg}$ parameter of the SSLCA and the number of spikes collected may be used to control the sparsity of the output.}

The performance seen on the $3\times 3$ and $8\times 8$ rescales are compared in \cref{fig:model:cifar-3v8}.  In both instances, the performance of the LCA is approached by the SSLCA.  However, the value of $Rf_{avg}$ that optimizes accuracy is not obvious based on the problem's statistics; using the dataset average works well for the $3\times 3$ case, but the $8\times 8$ case requires a smaller $Rf_{avg}$ in order to encourage more output activity, which translates into higher accuracy.  At values of $Rf_{avg}$ approaching the memristive device's minimum, the algorithm breaks down, as seen by the decreasing accuracy.  This result can be explained through \cref{eq:model:sslca-uninhib,alg:model:sslca}: small $Rf_{avg}$ results in a lower $Q_1$ and thus a smaller $C$, reducing the smoothing of input spike activity and in turn producing less consistent patterns of output spikes.

Another facet investigated was how different $K$ factors (spike duty cycles) affected the overall classification accuracy of the system.  The result is shown in \cref{fig:model:cifar:k-score}; generally, a higher input duty cycle $K_{in}$ performed better, and a lower output duty cycle $K_{out}$ performed better.  Intuitively this makes sense: larger duty cycles for input spikes means that more spikes are expected to work together when forming a single output spike; smaller duty cycles for output spikes means more time spent collecting input spikes, and thus each output spike represents a better average of the input spikes triggering it.

Device variability was also considered.  Previous work has demonstrated significant variance from one read to the next \cite{Degraeve2015}.  To test how the SSLCA performed with imperfect hardware, we implemented three types of conductance deviations: read deviation, write deviation using offline training, and write deviation with online training.  Read deviation was re-calculated after every output spike to better simulate the time-varying nature of read randomization, and varied the effective conductance of a device uniformly by $\pm$\SIrange{0}{80}{\percent} (a standard deviation of \SIrange{0}{46}{\percent}).  Write deviation with offline training consisted of training the model without variance, and then varying the conductance uniformly by $\pm$\SIrange{0}{180}{\percent} (not allowed to drop below \SI{0}{\siemens}; a standard deviation of \SIrange{0}{104}{\percent}).  Write deviation with online training was applied after each application of Oja's rule, and modified the target conductances uniformly by $\pm$\SIrange{0}{30}{\percent} (a standard deviation of \SIrange{0}{17}{\percent}).

These results can be seen in \cref{fig:model:cifar:variability}.  Neither read variability nor offline-trained write variability were found to have a significant impact.  For online training, write variability could be tolerated up to \SI{3}{\%}.  This result is satisfactory for 4-bit learning, as described in \mbox{\cref{sec:model:training}}.  Should better variability resistance be required, prior work on imperfect weight updates has indicated that sensitivity to these deviations might be further mitigated with a more aggressive training regimen that deliberately changes the magnitude of weight updates for greater effect \mbox{\cite{Woods2015tnano}}.

\fig[label=fig:model:cifar:variability]
    {\graphic{plots/variability-cifar-combine-score.pdf}}
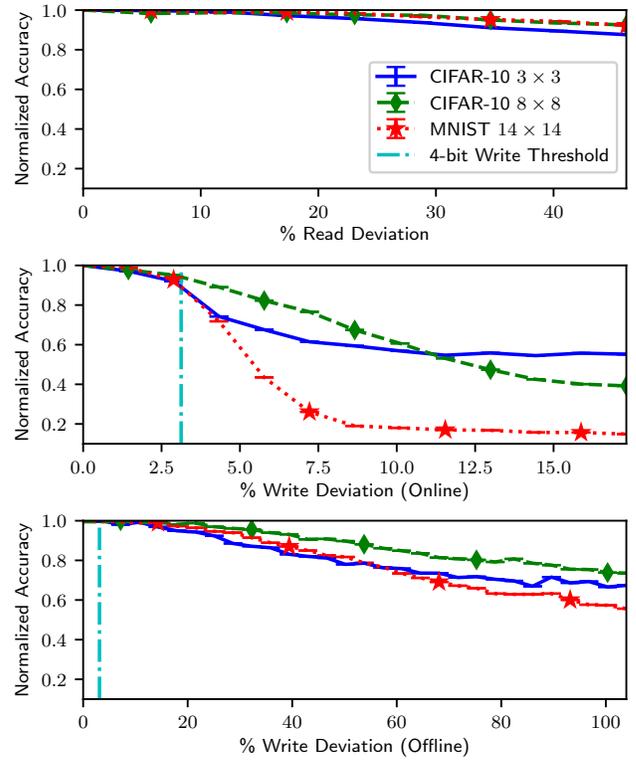
    {The effects on the SSLCA of conductance variability during each read cycle (the period of time between two output spikes), during each write cycle (online training), or when a weight matrix learned offline is written to the memristive crossbar (offline training).  Our results showed that unmitigated write deviations become serious for online algorithm stability after \SI{3}{\%}.  However, using offline training or modifying the training approach as previously reported helps significantly \cite{Woods2015tnano}.  For CIFAR-10 $3\times 3$, $8\times 8$, and MNIST $14\times 14$, $Rf_{avg} = 0.46, 0.425, 0.35$, respectively. Accuracies were normalized based on performance without deviations.}

\subsubsection{Power}\label{sec:model:cifar:power}
\fig[label=fig:model:cifar:power]
    {\graphic{plots/cifar-combine-power8x8.pdf}}
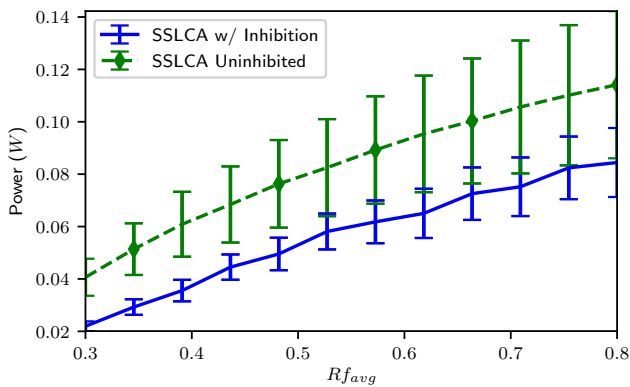
    {Power for $8\times 8$ CIFAR-10.  Inhibition produced lower, more consistent power consumption due to its suppression of input spikes.  Higher values of $Rf_{avg}$, which encourage the network to learn more conductive RFs, consumed more power accordingly.}

The inhibited SSLCA exhibited extremely low power consumption on the CIFAR-10 task scaled to $8\times 8$ with $128$ neurons ($2\times$ completion); at the optimal $Rf_{avg} = 0.43$, the consumption was just \ThisCifarEightPower\ (\cref{fig:model:cifar:power}) with a throughput of \ThisCifarEightThroughput.  Compared with prior work such as Knag \etal\ \cite{Knag2015}, whose lowest energy consumption was \KnagLowPower, this was a \ThisKnagCifarLowPower\ reduction in energy consumption for \ThisKnagCifarLowThroughput\ the throughput during inference \cite{Knag2015}.  At their high throughput (\SI{310}{\mega\hertz}), the SSLCA exhibited a \ThisKnagCifarHighPower\ reduction in energy consumption with a still substantially improved \ThisKnagCifarHighThroughput\ throughput.

Spiking architectures are often considered to produce power savings, though the extent of these savings has been a topic of discussion for some time \cite{Maass2015}; we investigated that claim in \cref{fig:model:cifar:k-all}.  Except for very large duty cycles, the spiking architecture's crossbar used less power than the non-spiking, voltage-scaled crossbar.  With a spiking implementation like the SSLCA, where input spikes are suppressed during an output spike, we would have expected the spiking to consume less power so long as $(1 - K_{out})K_{in} < Rf_{input}$.  This is a result of average power scaling with the square of voltage versus linearly with a duty cycle.  The SSLCA surpasses this expectation due to the additional input spike suppression implemented through the inhibition mechanism.  Interestingly, the standard deviation for the SSLCA's power was also substantially lower, probably as a result of columns in the SSLCA not being grounded, unlike the LCA, which sinks all current into a virtual ground \cite{Woods2015tnano,Woods2015}.

\subsubsection{Information Retention for Deep Learning}\label{sec:results:cifar:soa}

While an SLP might be used in practice due to the simplicity of its implementation, it does not adequately express the depth of the information contained in the input dataset.  To determine how much useful information was retained by both the LCA and SSLCA encodings, we used these architectures to encode augmented, full-size $32\times 32$ CIFAR-10 input images using convolutions of different sizes.  The convolved, sparse coded input images were then passed as input to a state-of-the-art deep learning architecture, the DenseNet-BC, presented by Huang \etal\ in 2016 \cite{Huang2016}.  This network architecture consists of a number of dense blocks that each halve the scale of the input data; within each dense block are many more layers, each accepting as input all previous layers within the dense block.  Using this setup with 3 dense blocks and parameters $L=190, k=40$, Huang \etal\ achieved \SI{96.54}{\%} accuracy on CIFAR-10 (with data augmentation) \cite{Huang2016}.  See Huang \etal\ for more further details on these parameters.

We tested our architecture by dividing the input CIFAR-10 image into non-overlapping patches of $S\times S$, and encoding each patch using either the LCA or the SSLCA.  For example, $S = 4$ implies that the $32\times 32$ CIFAR-10 image was broken into $8\times 8$ non-overlapping regions of $4\times 4$; each region was then sparse coded, and the resulting ``image'' consisting of all such encodings was passed to the DenseNet.  For the $2\times$ networks with $S = 4$, this means that rather than receiving each image as a $32\times 32\times 3$ spatial array, we passed in an $8\times 8\times 96$ spatial array.  For the $0.5\times$ networks, the spatial array passed would only have a depth of $24$.  The SSLCA was configured with $Rf_{avg} = 0.45$.

In order to allow each DenseNet a similar amount of expression for its classification, we parametrized the DenseNet so that the final dense block would output a $4\times 4$ spatial array; the original paper's final block output an $8\times 8$ array.  To accomplish this, each DenseNet had a number of dense blocks $B = -1 + \log_2 \frac{32}{S}$.  To hold the number of computations that each DenseNet performed roughly equivalent, we chose $L=40$, $k=12$, and the number of filters on the initial convolution before the first dense block was $k_0=6S^2B$ rather than $16$.  The limitation of this approach is that the number of tunable parameters becomes significantly larger with larger values of $S$, creating a greater potential for overfitting.

Rather than $300$ epochs with mini-batches of $64$ samples, we used $150$ epochs and mini-batches of $32$ samples to train these networks.  We trained using stochastic gradient descent, with an initial learning rate of $0.1$; after $75$ epochs this was reduced to $0.01$, and after $112$ epochs this was further reduced to $0.001$.  Simulations were done with keras; the DenseNet implementation can be found at \url{https://github.com/titu1994/DenseNet}, and keras can be found at \url{https://github.com/fchollet/keras}.  Each accuracy measurement was the result of a single trial, so some stochasticity is embedded in the reported results.  They are nonetheless internally consistent.  Results are shown in \cref{fig:model:cifar:soa,fig:model:cifar:soa-scale}.

\fig[label=fig:model:cifar:soa-scale]
    {\graphic{plots/cifar-soa-s.pdf}}
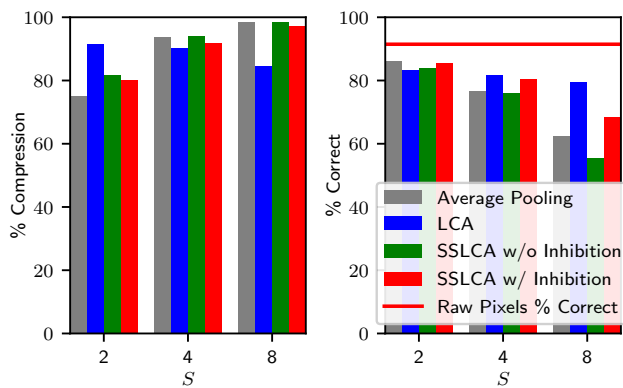
    {Comparison of each algorithm at different encoding scales $S$.  Inhibition always improved performance when using the deep classifier, without significantly affecting compression.  The LCA does not change significantly due to a fixed $\lambda$ threshold parameter; the SSLCA might achieve a similar effect by collecting more spikes, but this would slow the algorithm's throughput.}

On the raw CIFAR-10 data, these conditions produced a classification accuracy of \SI{92}{\%}.  With the analog LCA $S=4$, the DenseNet achieved a classification accuracy of \SI{82}{\%}, compressing the data down by \mbox{\SI{90}{\%}}.  The SSLCA produced an accuracy of \SI{80}{\%} with \mbox{\SI{92}{\%}} compression.  In the context of the deep classifier, we found a direct and inverse relationship between the LCA's RMSE and the classifier's accuracy (\mbox{\cref{fig:model:cifar:soa}}).  Compression was calculated as $1 - \#\ active\ neurons\times(\log_2(\#\ neurons) + 4)$ divided by the number of bits in the image ($8\times W\times H$).  This represents the minimum number of bits to send a neuron index and its 4-bit spike count per active neuron.

For different values of $S$, the LCA maintained similar compression factors, due to the threshold $\lambda$ being held constant with an increasing number of inputs, leading to more active outputs (\cref{fig:model:cifar:soa-scale}).  In contrast, the SSLCA's sparsity comes from the number of spikes collected, which was fixed at $10$ for all experiments.  Thus, for larger patch sizes, more and more sparse representations were created, resulting in lower accuracy but higher compression.  These parameters are all configurable and could be used to trade off between accuracy and compression, but these values were chosen as they produced roughly equivalent compression at $S=4$.  If the input dataset has greater covariance, then the accuracy loss would be lower for higher compression rates.  Inhibition was always beneficial with a deeper classifier, and each increase in accuracy aligned with lower RMSE without exception.  \mbox{\Cref{fig:model:cifar:soa-scale}} also includes compression factors and accuracies for downsampling the input image using average pooling across groups of $S\times S$ pixels.  This demonstrates a baseline image compression technique against the sparse coding achieved by the LCA family.  Generally, the LCA methods produced greater accuracies at comparable, high levels of compression than the average pooling method.


\subsection{MNIST}
The second dataset, MNIST, consisted of \SI{70000}{} $28\times 28$ grayscale images, each containing a single, centered, hand-written digit \cite{Lecun1998}.  Again for faster simulations, this dataset was scaled down to $14\times 14$.  The test set contains an equal number of each digit class, so a simple accuracy was tabulated for each algorithm.

\fig[label=fig:model:mnist:score-power]
    {\graphic{plots/mnist-rf-score.pdf}}
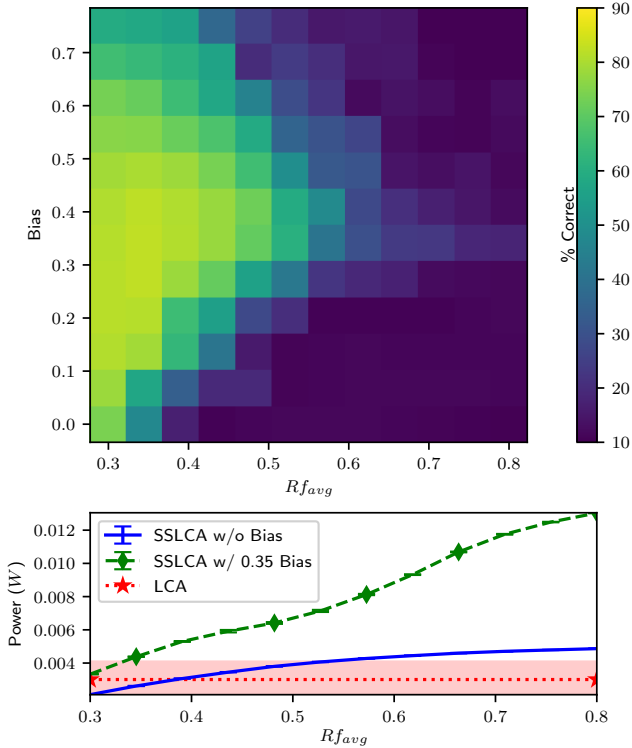
    {Results of MNIST scaled to $14\times 14$.  Since MNIST has a much lower average input value than CIFAR, a bias needed to be applied to compensate for the additional current lost from the neurons back into the crossbar (\cref{eq:model:sslca-charge}).  Since a bias also increased the duty cycle of each input signal, more power was consumed.}

\subsubsection{Accuracy}
The SSLCA as defined up to this point performed notably worse on MNIST than the non-spiking LCA (\cref{fig:model:mnist:score-power}).  Unlike CIFAR, which has an average input value of $0.47$, MNIST has an average input value of only $0.13$.  Since the SSLCA was designed deliberately to include a leak current through the crossbar, this lower input intensity could not sustain neuron charge reliably, leading to more random patterns of input events being encoded in each output spike.  We found that applying a bias signal, by redefining the duty cycle of each input signal from $K_{max}k_{input}$ to $K_{max}(bias + (1 - bias)k_{input})$, we could remedy this problem while preserving the power gains of the SSLCA architecture.  For MNIST, we found that a bias of $0.35$ boosted performance from \SI{77}{\percent} correct classification up to \SI{84}{\percent}, versus a performance of \SI{88}{\percent} by an optimal, analog LCA.  The MNIST experiments' responses to write deviations were not found to be significantly different than the CIFAR experiments', shown in \cref{fig:model:cifar:variability}.

\subsubsection{Power}
The power savings on MNIST, even with the bias, were in-line with those found for CIFAR: \ThisMnistPower\ for \ThisMnistThroughput.  Note that the increased power savings compared to CIFAR (which consumed \ThisCifarEightPower) were due to the lower relative number of outputs to inputs: additional neurons are more expensive than additional input lines (partially due to the comparator, though mostly due to the crossbar).  While the cost of additional inputs is different from the cost of additional columns, the SSLCA still demonstrates \bigOn\ scaling in both dimensions.

As seen in \cref{fig:model:mnist:score-power}, a non-spiking approach might consume less power on MNIST due to the low $Rf_{avg}$ of the dataset.  On the other hand, the power presented for the LCA does not include inhibitory logic, unlike the SSLCA: it would be difficult to include inhibition logic without closing the already-narrow margin.

\section{Conclusion}\label{sec:conc}
Our work demonstrated that memristive devices with a low conductance ratio could be used in the design of a fast, low-power sparse coding circuit, with in-situ learning, as long as their conductances could be set within \SI{\pm 3}{\percent}.  This requirement matches the 4-bit resolution required by other neuromorphic architectures.  Our proposed circuit was both fast and energy-efficient, improving upon a previously published all-CMOS ASIC with \ThisKnagCifarHighThroughput\ the throughput while using \ThisKnagCifarHighPower\ less energy per input.  The resulting sparse codes were also shown to be of a high quality.  When evaluated with a state-of-the-art deep learning network, our circuit demonstrated a reduction in relative accuracy of only \SI{2.4}{\%} (to \SI{80}{\%} from \SI{82}{\%}) compared to an optimal, analog sparse coding algorithm.  Our circuit maintained this fidelity while compressing the input data by \mbox{\SI{92}{\%}}.  These figures are all affected by circuit parameters that could be adjusted for higher accuracy and lower compression.  We showed that even datasets with low input activity, such as MNIST, could be properly represented through the use of a bias.  The proposed SSLCA architecture was demonstrated to be very resistant to device variations, particularly when used with offline training.  Sparse coding algorithms such as the SSLCA could be used to greatly reduce communication bandwidth between visual sensors and other processing algorithms, such as deep-learning networks.

\section*{Acknowledgement}
The authors would like to thank Garrett Kenyon of Los Alamos National Laboratories for helpful discussions concerning the LCA.  This work was supported by the National Science Foundation under award \# 1028378 and by DARPA under award \# HR0011-13-2-0015.  The views expressed are those of the author(s) and do not reflect the official policy or position of the Department of Defense or the U.S. Government.  Approved for public release, distribution is unlimited.

\bibliographystyle{./sty/ieee/IEEEtran-nomonth}
\bibliography{./sty/ieee/IEEEabrv,biblio}

\newif\ifsupplementaltoo
\supplementaltootrue

\ifsupplementaltoo
\clearpage
\renewcommand{\appendixname}{Supplementary Material}
\appendix
\setcounter{figure}{0}
\makeatletter
\renewcommand{\thefigure}{S\@arabic\c@figure}
\makeatother
\fig[label=fig:model:cifar:k-score, pos=H]
    {\graphic{plots/cifar-k8x8-score.pdf}}
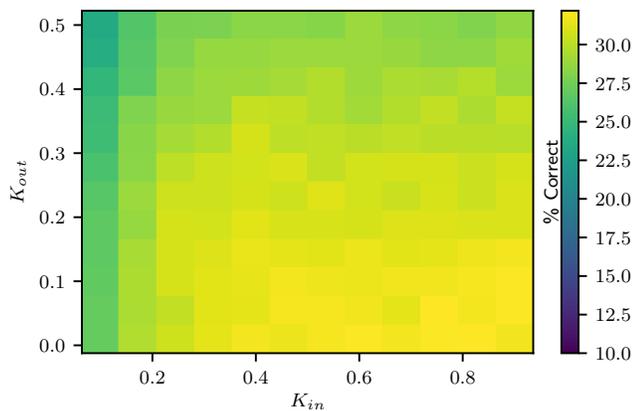
    {CIFAR-10 spike accuracy with different spike duty cycles.  High duty cycles for input spikes and low duty cycles for output spikes are the most accurate, but require more power (\cref{fig:model:cifar:k-all}).}

\fig[label=fig:model:cifar:k-all, pos=H]
    {\graphic{plots/cifar-k8x8-all.pdf}}
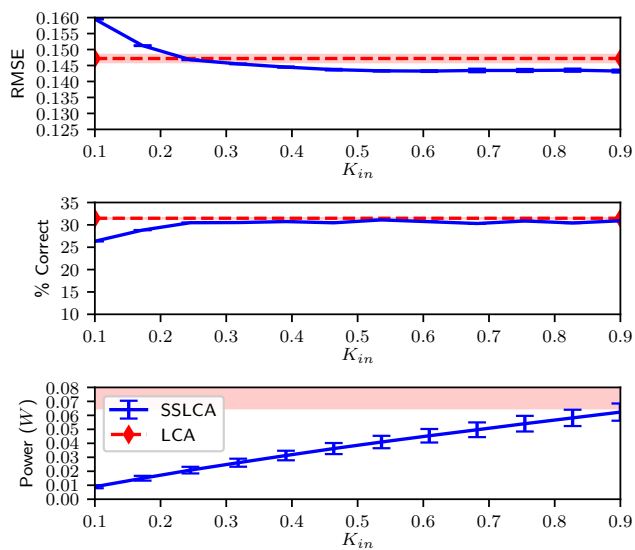
    {Power and accuracy trade-offs on CIFAR-10 with varied $K_{in}$.  LCA power shown represents only a voltage-scaled crossbar, to directly compare spiking and non-spiking approaches.  Spiking always consumed less power than a voltage-scaling approach, a combination of the dataset having a high average input and the spiking algorithm utilizing inhibition of input signals (see \cref{fig:model:cifar:power} for the effects of inhibition on power consumption).}

\fig[label=fig:model:cifar:soa, pos=H]
    {\graphic{plots/cifar-soa-combine.pdf}}
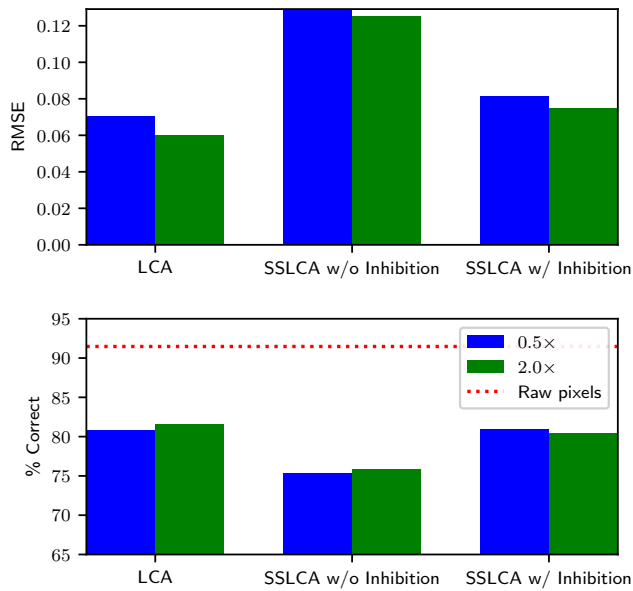
    {Comparison of different algorithms and different completenesses at $S=4$.  The SSLCA is capable of matching the accuracy of the LCA when a deeper classifier is used.}

\fi
\end{document}